\documentclass{article}

\usepackage[final]{neurips_2019}

\usepackage[utf8]{inputenc} 
\usepackage[T1]{fontenc}    
\usepackage[hidelinks]{hyperref}       
\usepackage{url}            
\usepackage{booktabs}       
\usepackage{amsfonts}       
\usepackage{nicefrac}       
\usepackage{microtype}      

\usepackage{ifthen}
\usepackage[export]{adjustbox}
\usepackage{amsmath} 
\usepackage{amssymb} 
\usepackage{graphicx}
\usepackage{pgf}

\DeclareMathOperator*{\argmin}{argmin}
\usepackage{tikz}
\usetikzlibrary{positioning}
\usepackage{longtable}
\usepackage{rotating}
\usepackage{array}
\usepackage{multicol}
\usepackage{multirow}

\renewcommand\vec{\mathbf}

\title{Image Quality Assessment for Rigid Motion Compensation}

\author{%
  Alexander Preuhs$^1$, Michael Manhart$^2$, Philipp Roser$^1$, Bernhard Stimpel$^1$,\\ \bf Christopher Syben$^1$, Marios Psychogios$^3$, Markus Kowarschik$^2$, Andreas Maier$^1$\\
  $^1$Pattern Recognition Lab, Friedrich-Alexander Universität, Erlangen, Germany\\
  $^2$Siemens Healthcare GmbH, Forchheim, Germany\\
  $^3$Department of Neuroradiology, Universitätsspital Basel, Basel, Switzerland \\
  \texttt{alexander.preuhs@fau.de} \\
}

\usepackage{xcolor}

\definecolor{purple}{RGB}{255,0,144}
\definecolor{FAUblue1}{RGB}{0,56,101}
\definecolor{FAUblue2}{RGB}{144,167,198}
\definecolor{FAUblue3}{RGB}{221,229,240}

\definecolor{FAUgray1}{RGB}{152,164,174}
\definecolor{FAUgray2}{RGB}{210,213,215}
\definecolor{FAUgray3}{RGB}{235,236,238}

\definecolor{custompurple}{RGB}{0,255,255}
\usepackage{tikz}
\usetikzlibrary{calc} 
\usetikzlibrary{positioning}
\usetikzlibrary{intersections} 

\usetikzlibrary{arrows, shapes, positioning, calc, automata}
\newcommand{\MarkRightAngle}[4][.6cm]
{\coordinate (tempa) at ($(#3)!#1!(#2)$);
	\coordinate (tempb) at ($(#3)!#1!(#4)$);
	\coordinate (tempc) at ($(tempa)!0.5!(tempb)$);
	\draw (tempa) -- ($(#3)!2!(tempc)$) -- (tempb);
}

\usetikzlibrary{decorations.text}

\begin{document}

\maketitle

\begin{abstract}
 Diagnostic stroke imaging with C-arm \textit{cone-beam computed tomography} (CBCT) enables reduction of time-to-therapy for endovascular procedures. 
 However, the prolonged acquisition time compared to helical CT increases the likelihood of rigid patient motion. 
 Rigid motion corrupts the geometry alignment assumed during reconstruction, resulting in image blurring or streaking artifacts. 
 To reestablish the geometry, we estimate the motion trajectory by an autofocus method guided by a neural network, which was trained to regress the reprojection error, based on the image information of a reconstructed slice. 
 The network was trained with CBCT scans from 19 patients and evaluated using an additional test patient. 
 It adapts well to unseen motion amplitudes and achieves superior results in a motion estimation benchmark compared to the commonly used entropy-based method.
\end{abstract}

\section{Introduction}
Mechanical thrombectomy \cite{powers20152015,berkhemer2015randomized}  is guided by an interventional C-arm system capable of 3-D imaging. Although its soft tissue contrast is comparable to helical CT scans, the prolonged acquisition time pose C-arm CBCT more susceptible to rigid head motion artifacts \cite{leyhe2017latest}. In the clinical workflow, however, it is desirable to reduce the time-to-therapy by avoiding prior patient transfers to helical CT or MR scanners \cite{psychogios2017one}. To this end, robust motion compensation methods are desirable.

Methods for rigid motion compensation can be clustered in four categories: 1) image-based autofocus \cite{sisniega2017motion,wicklein2012image}, 2) registration-based \cite{Ouadah2016}, 3) consistency-based \cite{Frysch2015,preuhs2018double,preuhs2019symmetry} and 4) data-driven \cite{bier2018detecting,latif2018automating,kustner2019retrospective}. 

Recent data-driven approaches use image-to-image translation methods based on GANs \cite{latif2018automating,kustner2019retrospective} or aim to estimate anatomical landmarks in order to minimize a \textit{reprojection error} (RPE) \cite{bier2018detecting}. 
The latter approach does not provide the required accuracy, whereas GAN-based approach are deceptive for clinical applications, as the data-integrity cannot be assured \cite{huang2019data}.

We propose a learning-based approach for rigid motion compensation ensuring data integrity. 
An image-based autofocus method is introduced, where a regression network predicts the RPE directly from reconstructed slice images. 
The motion parameters are found by iteratively  minimizing the predicted RPE using the Nelder-Mead simplex method \cite{olsson1975nelder}. 

\section{Motion Estimation and Compensation Framework}
\paragraph{Autofocus Framework:}
Rigid motion is compensated by estimating a motion trajectory $\mathcal{M}$ which samples the motion at each of the $N$ acquired views within the trajectory \cite{Kim2014}. $\mathcal{M}$ contains the motion matrices $\vec{M}_i$, where each motion matrix $\vec{M}_i \in \mathbb{SE}(3)$~---~with $\mathbb{SE}(3)$ being the special Euclidean group~---~describes the patient movement at view $i \in [1,N]$. 
The motion matrices can be incorporated in the backprojection operator of a \textit{filtered backprojection}-type (FBP) reconstruction algorithm. 
We denote the reconstructed image in dependence of the motion trajectory by FBP$_y(\mathcal{M})$, where FBP$_y$ is the FDK-reconstruction \cite{feldkamp1984practical} from projection data $y$. 
In the following, FBP$_y$ will reconstruct the central slice on a $512^2$ pixel grid using a sharp filter kernel to emphasize motion artifacts.

Typical autofocus frameworks (cf.~\cite{sisniega2017motion}) estimate the motion trajectory based on an \textit{image quality metric} (IQM) evaluated on the reconstructed image by minimizing
\begin{equation}
\argmin_\mathcal{M}   \enspace\text{IQM}(\text{FBP}_y(\mathcal{M})) \enspace.
\label{eq:IQM}
\end{equation}
A common problem in solving \eqref{eq:IQM} is the non-convexity of the IQM, which is typically chosen to be the image histogram entropy or total variation of the reconstructed slice. To overcome this limitation, we propose to replace the IQM by a network architecture that is trained to regress the RPE, which was shown to be quasi convex for a geometric reconstruction problem \cite{ke2007quasiconvex}. 
\paragraph{Learning to Assess Image Quality:}
\label{sec:regression}
Let $\mathcal{X}$ be a set of \mbox{3-D} points $\vec{x} \in \mathbb{P}^3$  uniformly sampled from a sphere surface and let the acquisition trajectory associated to a dataset $y$ be defined by projection matrices $\vec{P}_i \in \mathbb{R}^{3\times4}$ mapping world points on the detector of a CBCT system at view $i$ \cite{hartley2003multiple}, then the RPE is computed as 
\begin{equation}
\text{RPE}(\mathcal{M}) =\frac{1}{|\mathcal{X}|N} \sum_{\vec{x} \in \mathcal{X},i\in N}||\vec{P_i}\vec{M_i}\vec{x} - \vec{P_i}\vec{x}||_2^2 \enspace.
\label{eq:rpe}
\end{equation}
This metric measures the reconstruction-relevant deviations induced by motion  \cite{strobel2003improving} and can thus be expected to be estimated directly from the reconstruction images. 
To this end, we devise a regression network learning the RPE directly from a reconstructed image. Our regression network $R$ consists of a feature extraction stage, pretrained on ImageNet and realized by the first 33 layers from a residual network, \cite{he2016deep} followed by a densely connected layer defining the regression stage. The cost function $L$ of the network is defined by the difference between the network-predicted RPE from a reconstruction slice with simulated motion trajectory $\mathcal{M}$ and the corresponding RPE as defined by Eq. \eqref{eq:rpe}
\begin{equation}
	L = || R(\text{FBP}_y(\mathcal{M})) - \text{RPE}(\mathcal{M})||_2^2 \enspace .
\end{equation}
For training, the projection data $y$ is ensured to be motion free, such that motion artifacts solely source from the virtual motion trajectory $\mathcal{M}$.
For training and testing, we use CBCT acquisitions (Artis Q, Siemens Healthcare GmbH, Germany) of the head ($N=496$) acquired from 20 patients which were split in 16 for training 3 for validation and 1 for testing. 
For each patient we simulate 450 random motion trajectories resulting in a training set of 7200 reconstructions.

\section{Experiments and Results}
For motion generation, we use rotational movements along the patient's longitudinal axis. The motion trajectory is modeled by an Akima spline \cite{akima1970new} with 15 equally distributed nodes inducing RPEs ranging from $0$\,mm to $0.6$\,mm. With the RPE measurement being sensitive to constant offsets, not inducing motion artifacts, we further only use motions affecting a third of the acquisition.

\begin{figure}
	\adjustbox{width=0.94\textwidth,trim=0ex 0pt 0ex 0pt}{
		\begin{tikzpicture}
		\node[inner sep=0pt, anchor = north west] (recos) at (-1,0)
		{
			\resizebox{!}{0.1\textheight}{
				\begin{tabular}{lr}
				\includegraphics[width=0.2\linewidth]{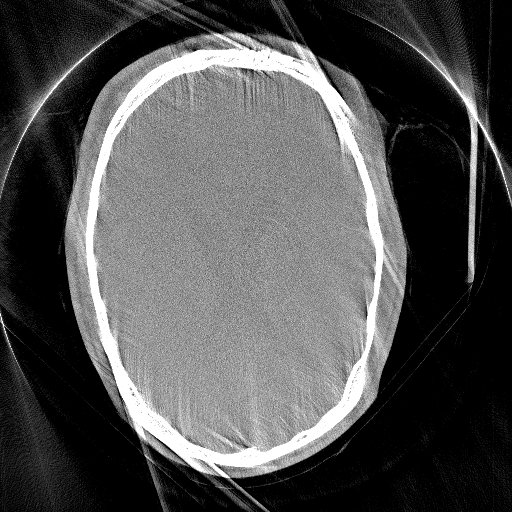}\\ \includegraphics[width=0.2\linewidth]{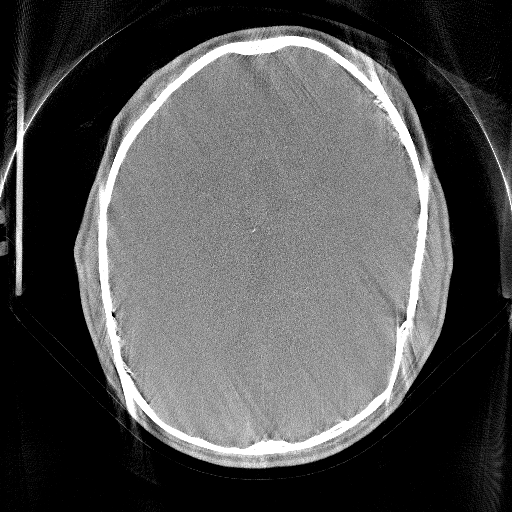}\\ \includegraphics[width=0.2\linewidth]{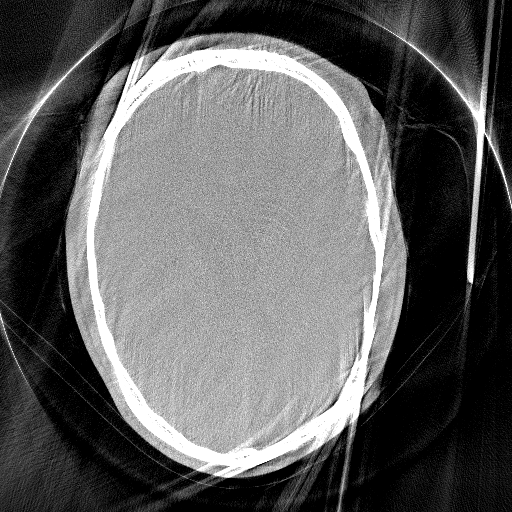}  
				\end{tabular} }};
		\node[inner sep=0pt, anchor = north east] (plot) at (-1,0.107)
		{\input{tikz/regression.pgf}};
		\node [inner sep=0pt] (A) at (-4.4,-1.6) {};
		\node [inner sep=0pt] (B) at (-4.4,-1.5) {};
		\node [inner sep=0pt] (C) at (-4.4,-1) {};
		\node [inner sep=0pt] (A1) at (-0.8,-0.65) {};
		\node [inner sep=0pt] (B1) at (-0.8,-2.2) {};
		\node [inner sep=0pt] (C1) at (-0.8,-3.8) {};
		\draw [-] (A) -- (C1);
		\draw [-] (B) -- (B1);
		\draw [-] (C) -- (A1);
		\end{tikzpicture}
	}
	\caption{Network estimated RPE and different reconstructions, all revealing a RPE of $\approx 0.34$\,mm.}
	\label{fig:test_accuracy}
\end{figure}

First, we inspect how well the network is able to regress the RPE on test and validation data. Then, in in an inverse crime scenario~---~i.e. the modeling capacity of the spline used for motion generation is equal to the spline used for motion compensation~---~we inspect the behavior for motion types significantly varying in their shape from any motion seen during training. In a last experiment we compare the performance of the network with a state-of-the-art IQM utilizing the histogram entropy. Therefore, we  deploy an inverse crime scenario and a more realistic case where we use 10 spline nodes for motion generation and 20 nodes for compensation.
\paragraph{Regression Network:}
 We use Adam optimization with learning rate of $0.00001$ and select the network parameters that achieved the best RPE prediction on our validation dataset. Our network achieves an average RPE deviation from the Gt of \mbox{$0.031$\,mm} on the test dataset, as depicted in Fig.\,\ref{fig:test_accuracy}.

\paragraph{Network Inference for Motion Compensation:}
\begin{figure}
	\adjustbox{width=0.94\textwidth,trim=0ex 0pt 0ex 0pt}{
		\input{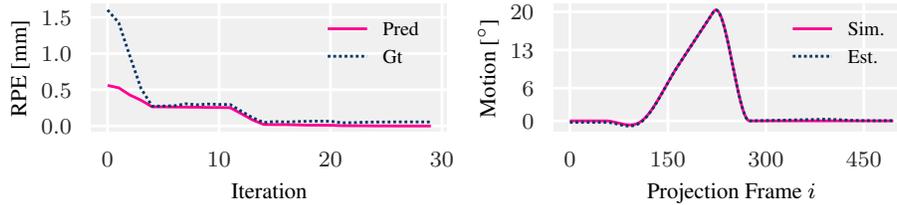}
	}
	\caption{Left: Network-predicted and Gt RPE in each iteration step of the optimization. Right: Simulated motion trajectory and estimated motion trajectory after optimization.}
	\label{fig:ic}
\end{figure}
Using the test patient, the network behavior for motion exceeding the RPE of the training process is inspected in an inverse crime scenario. The simulated motion trajectory is depicted in Fig.\,\ref{fig:ic} together with the estimated motion trajectory after optimization using the network as IQM (cf.\,Eq.\,\ref{eq:IQM}). For each iteration of the optimization process the network predicted RPE together with the corresponding Gt RPE is depicted. While the RPE is underestimated within the first iterations, the proportionality is still kept, guiding the optimization to a motion free reconstruction.

\begin{figure}
	\begin{center}
		\resizebox{1\textwidth}{!}{
			{\def\arraystretch{1}\tabcolsep=2pt 
				\begin{tabular}{cccccccc}
					\begin{tikzpicture}
					\node[anchor=south west,inner sep=0,outer sep=0] (image) at (0,0) {%
						\includegraphics[width=0.2\textwidth]{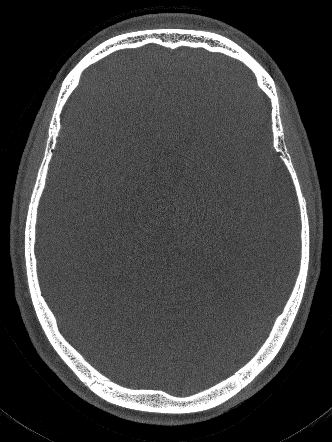}%
					};
					\begin{scope}[x={(image.south east)},y={(image.north west)}]
					\node[white,anchor=west] at (-0.0,0.9) {Gt}; 
					\end{scope}
					\end{tikzpicture}
					&
					\begin{tikzpicture}
					\node[anchor=south west,inner sep=0,outer sep=0] (image) at (0,0) {%
						\includegraphics[width=0.2\textwidth]{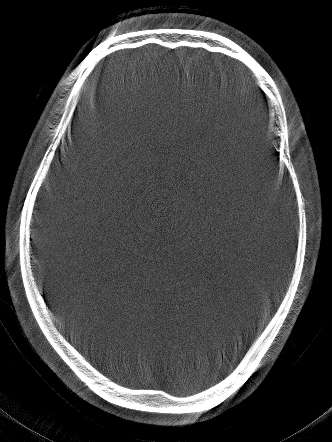}
					};
					\begin{scope}[x={(image.south east)},y={(image.north west)}]
					\node[white,anchor=west] at (-0.0,0.9) {Mo}; 
					\end{scope}
					\end{tikzpicture}
					&{\color{white} a}&
					\begin{tikzpicture}
					\node[anchor=south west,inner sep=0,outer sep=0] (image) at (0,0) {%
						\includegraphics[width=0.2\textwidth]{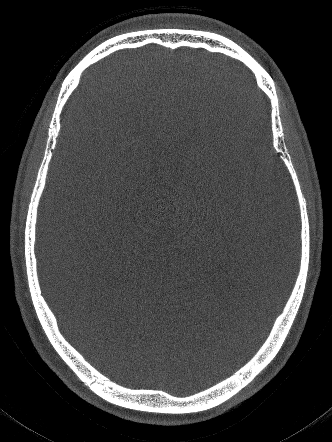}%
					};
					\begin{scope}[x={(image.south east)},y={(image.north west)}]
					\node[white,anchor=west] at (-0.0,0.9) {Ent}; 
					\end{scope}
					\end{tikzpicture}
					&
					\begin{tikzpicture}
					\node[anchor=south west,inner sep=0,outer sep=0] (image) at (0,0) {%
						\includegraphics[width=0.2\textwidth]{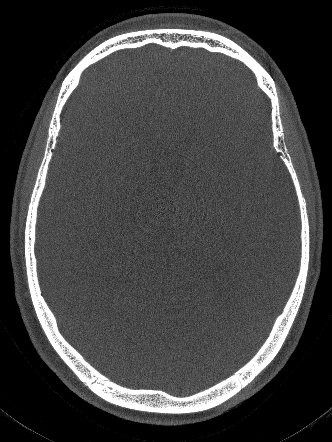}%
					};
					\begin{scope}[x={(image.south east)},y={(image.north west)}]
					\node[white,anchor=west] at (-0.0,0.9) {Pro}; 
					\end{scope}
					\end{tikzpicture}
					&{\color{white} a}&
					\begin{tikzpicture}
					\node[anchor=south west,inner sep=0,outer sep=0] (image) at (0,0) {%
						\includegraphics[width=0.2\textwidth]{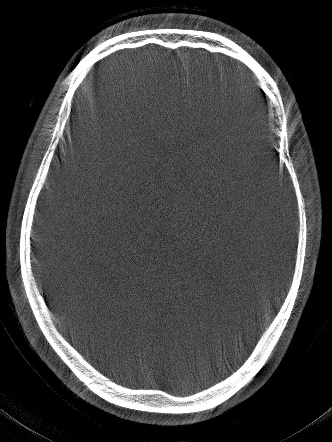}%
					};
					\begin{scope}[x={(image.south east)},y={(image.north west)}]
					\node[white,anchor=west] at (-0.0,0.9) {Ent}; 
					\end{scope}
					\end{tikzpicture}
					&
					\begin{tikzpicture}
					\node[anchor=south west,inner sep=0,outer sep=0] (image) at (0,0) {%
						\includegraphics[width=0.2\textwidth]{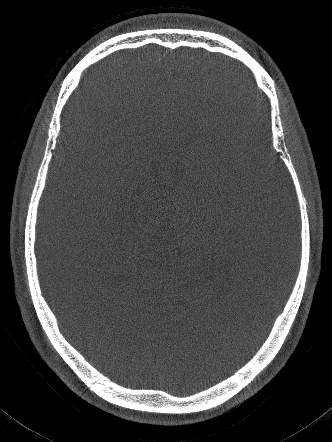}
					};
					\begin{scope}[x={(image.south east)},y={(image.north west)}]
					\node[white,anchor=west] at (-0.0,0.9) {Pro}; 
					\end{scope}
					\end{tikzpicture}				
					\\
					\multicolumn{2}{c}{Ground Truth (Gt) and Motion Affected}&&
					\multicolumn{2}{c}{Inverse Crime Compensation}&&
					\multicolumn{2}{c}{Clinical Setting (Entropy and Proposed)}
				\end{tabular}%
		}}
	\end{center}
	\caption{Reconstructions of the test patient using [500-2000] HU window. In the inverse crime scenario, the SSIM to the Gt is $0.84$ (Ent/Gt) and $0.95$ (Pro/Gt), respectively for the entropy (Ent) and proposed (Pro) measure. For the more realistic setting (Clinical Setting) the SSIM is $0.65$ (Ent/Gt) and $0.84$ (Pro/Gt), respectively.
	}
	\label{fig:motion_reco}
\end{figure}
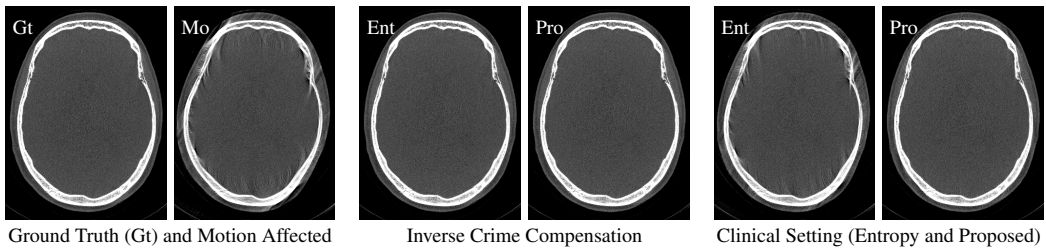

Figure \ref{fig:motion_reco} compares the proposed network-based IQM with the entropy-based IQM. The optimization process is identically for both metrics. In an inverse crime scenario both methods can restore the original image quality, however, in a more realistic setting the image entropy is stuck in a local minimum, whereas the network is able to lead the optimization to a nearby motion-free solution.

\section{Conclusion and Discussion}
We present a novel data driven autofocus approach lead by a convolutional neural network. 
The network is trained to predict the RPE given a slice of a CBCT reconstruction. The final motion compensated reconstruction is solely based on the projection raw-data and the estimated motion trajectory. 
This allows us to devise a learning-based motion compensation approach while ensuring data integrity. We showed that  the network is capable of generalizing well to unseen motion shapes and achieves higher SSIM compared to a state-of-the-art IQM measure.  

\textbf{Disclaimer:} The concepts and information presented in this paper are based on
research and are not commercially available.

\medskip
\small
\bibliographystyle{splncs04}
\bibliography{main}

\end{document}